\documentclass[msmath,amssymb,aps,pra,twocolumn,epsfig,showpacs,bibliography,lengthcheck,superscriptaddress]{revtex4-2}

\usepackage{graphicx}
\usepackage{dcolumn}
\usepackage{bm}
\usepackage{hyperref}
\usepackage{epstopdf}
\usepackage{subcaption}
\usepackage{caption}
\usepackage{footmisc}
\usepackage{amsmath}
\usepackage{color}
\usepackage{ulem}

\begin{document}
\title{Suppression of heading errors in Bell-Bloom optically pumped  free-induction-decay alkali-metal atomic magnetometers}%
\author{S.-Q. Liu}
\affiliation{Department of Precision Machinery and Precision Instrumentation, Key Laboratory of Precision Scientific Instrumentation of Anhui Higher Education Institutes, University of Science and Technology of China, Hefei 230027, China}

\author{X.-K. Wang}
\affiliation{Department of Precision Machinery and Precision Instrumentation, Key Laboratory of Precision Scientific Instrumentation of Anhui Higher Education Institutes, University of Science and Technology of China, Hefei 230027, China}

\author{X.-D. Zhang}
\affiliation{College of Physics and Optoelectronic Engineering, Shenzhen University, Shenzhen 518060, China}
\affiliation{Beijing Computational Science Research Center, Beijing 100193, China}

\author{W. Xiao}
\email{xiao$_$wei@pku.edu.cn}
\affiliation{State Key Laboratory of Advanced Optical Communication Systems and Networks, School of Electronics, and Center for Quantum Information Technology, Peking University, Beijing 100871, China}
\affiliation{MIIT Key Laboratory of Complex-field Intelligent Sensing, Advanced Research Institute of Multidisciplinary Science, Beijing Institute of Technology, Beijing 100081, China}

\author{D. Sheng}
\email{dsheng@ustc.edu.cn}
\affiliation{Department of Precision Machinery and Precision Instrumentation, Key Laboratory of Precision Scientific Instrumentation of Anhui Higher Education Institutes, University of Science and Technology of China, Hefei 230027, China}
\affiliation{Hefei National Laboratory, University of Science and Technology of China, Hefei 230088, China}

\begin{abstract}
Heading errors of atomic magnetometers refer to the dependence of measurement results on the sensor orientation with respect to the external magnetic field. There are three main sources of such errors: the light shift effect, the linear nuclear-spin Zeeman effect, and the nonlinear Zeeman effect. In this work, we suppress the former two effects by using the Bell-Bloom optical pumping method and probe the atomic signals while the pumping beam is off, and focus on the heading error induced by nonlinear Zeeman effect while the sensor operates in the geomagnetic field range. We demonstrate several schemes to suppress this remaining heading error within 1 nT using a single magnetometer or a comagnetometer. In the magnetometer system, two schemes are developed to average out the horizontal atomic polarization in space or in time, respectively. In the comagnetometer system, we combine the simultaneously measured Larmor frequencies of two different kinds  of alkali atoms to either suppress the heading error or extract the orientation of the pumping beam relative to the bias field.
\end{abstract}

\maketitle
\section{Introduction}
Atomic scalar magnetometers directly measure the magnitude of the bias magnetic field using the Zeeman effect. For sensors based on alkali vapors, the sensor noise has reached fT/Hz$^{1/2}$ level~\cite{sheng2013,Lucivero2022}. Together with the intrinsic stability of atom transitions, these sensors are widely applied in different research fields. For example, atomic scalar magnetometers have been used to calibrate fluxgates in space science~\cite{acuna2002,patton2014}, enable out-of-shield biomagnetic field detections~\cite{limes2020,xiao2023}, monitor the bias field fluctuations on the pT level for precision measurements~\cite{abel2020,rosner2022}, and serve as the fundamental units in comagnetometers based on comparing atomic spin precession frequencies~\cite{kimball2017,wang2020}. However, an important factor to limit the accuracy of the scalar magnetometer readout is its heading error, where the measurement results correlate with the relative angle between the sensor and the bias field. It is important to both identify and suppress such an error in practical applications. 

A common source of the heading error is the light shift effect~\cite{Oelsner2019,yu2022}, which is from either real or virtual atomic transitions~\cite{barrat1961a,barrat1961} due to atom-photon interactions. Light shifts due to virtual transitions can be decomposed to several components according to symmetries~\cite{happer1967}, among which the vector part~\cite{mathur1968} is the most important one in experiments conducted in atomic cells filled with quenching gases. The vector light shift is due to the interaction between atoms with the circular polarization component of a detuned beam, and its effect is equivalent to an effective magnetic field along the beam propagation direction. Experimental methods to circumvent this problem include locking beams on resonance, employing atomic-alignment-based magnetometry using only linearly polarized beams~\cite{abel2020,zhangrui2023}, and averaging the beam polarization out in time~\cite{benkish2010} or space~\cite{yu2022}. In the past decade, all-optical free-induction-decay (FID) magnetometers were developed~\cite{grujic2015,hunter2018,limes2020}, where the pumping and probe stages are separated in time so that the magnetometer works in the pulse mode and the light shift from pumping beams is absent when the signal is recorded.

The other main source of the heading error is due to the non-zero nuclear spin. For alkali atoms, the linear Zeeman effect (LZE) of the nuclear spin makes a slight difference in the absolute value of atomic gyromagnetic ratios for atoms in different ground hyperfine states. This effect can induce heading errors if the sensor signal has contributions from both hyperfine states, and the ratio of the two contributions changes with the sensor orientation. This is a serious problem for certain classes of magnetometers~\cite{Hewatt2024}. For FID magnetometers, an experiment scheme based on dual orthogonal probe beams is developed to suppress this effect~\cite{lee2021} by taking advantage of the opposite signs of atomic gyromagnetic ratios in the two ground hyperfine states. Another method based on a double-frequency fitting function is also proposed to solve this problem~\cite{Hewatt2024}. A more direct way to suppress this effect is to selectively polarize atoms onto the upper hyperfine state. This can be achieved using Bell-Bloom optical pumping method~\cite{bell1961}, which is due to the facts that the Zeeman transition line width is relatively narrow compared with the separation between resonant frequencies of the two ground states, and only the upper hyperfine state has a dark state in optical pumping processes.

Even alkali atoms are all pumped to the upper hyperfine state, the nonlinear Zeeman effect (NLZE) can still lead to an additional energy shift that is dependent on the specific Zeeman sublevel. In this way, the dependence of the measurement results on the sensor orientation remains.  {While experiment schemes to directly cancel the NLZE by light shifts~\cite{jensen2009} have been developed, the main strategy in the literature to suppress NLZE} is based on symmetry considerations so that the net atomic orientation in the longitudinal direction is nulled. This has been implemented by either modulating the atomic polarization using electric-optical modulators~\cite{benkish2010} or making measurements based on atomic alignment~\cite{zhangrui2023}. Recently, a systematic study of this effect in FID magnetometers has been performed by Romalis' group~\cite{lee2021}, and a new method is brought up in that work, which extracts the sensor orientation from the ratio between the longitudinal and transverse atomic polarization, and corrects the NLZE effect according to the derived formula. 

In this work, we study the heading error of a compact Rb FID magnetometer, where the signal is based on atomic orientations directly generated by an amplitude-modulated pumping beam. Besides identifying sources of the measured heading error, we also {develop generic} methods in two different systems to suppress the dominant NLZE-induced heading error, {which is controlled to be below 1 nT when the bias field is in the geomagnetic field range}. Following this introduction, Sec.~II introduces theoretical background of this work, Sec.~III describes the sensor setup and measurement scheme, Sec.~IV focuses on the sources of heading errors and experiment schemes to suppress the main source, and Sec.~V concludes the paper.

\section{Theoretical Background}
In this section and the rest of the paper, we are considering a general case that the bias field is along the $z$ direction, and the pumping beam is aligned in the $xz$ plane with an angle $\theta$ relative to the bias field direction.
\subsection{Bell-Bloom optical pumping}
As first pointed out by Bell and Bloom~\cite{bell1961}, transverse atomic polarization can be directly generated by the synchronized optical pumping method. In this work, we focus on pumping beams that are turned on and off by acousto-optical modulators (AOMs). The {electron spin} polarization $\bm{P}$ can be described by the Bloch equation
\begin{equation}
\frac{d\bm{P}}{dt}=-\gamma\mathbf{B}\times\bm{P}+R_{op}(t)(\bm{s}-\bm{P})-\Gamma_{1,0}P_z\hat{z}-\Gamma_{2,0}(P_{x}\hat{x}+P_y\hat{y}),
\end{equation}  
where $\bm{s}$ is the pumping beam photon spin~{\cite{appelt98}}, $\Gamma_{1(2),0}$ is the longitudinal (transverse) depolarization rate in the absence of light, $R_{op}(t)$ is the real-time optical pumping rate with its time-averaged value as $\langle R_{op}\rangle$, and $\gamma$ is the atomic gyromagnetic ratio.

In the large magnetic field limit, $\gamma B\gg \langle R_{op}\rangle$ and {$\Gamma_{1(2),0}$}, the steady state solution for longitudinal polarization $P_l$ and transverse polarization $P_{t}$ can be expressed as:
\begin{eqnarray}
P_l&=&\frac{\langle R_{op}\rangle\cos\theta}{\Gamma_{1,0}+\langle R_{op}\rangle},~\label{eq:Pz}\\
P_t&=&\frac{\langle R_{op}\rangle\sin\theta}{\Gamma_{2,0}+\langle R_{op}\rangle}~\label{eq:Pt}.
\end{eqnarray}
Generally speaking, $\Gamma_{1,0}$ is not equal to $\Gamma_{2,0}$ due to extra contribution to transverse depolarization such as the spin-exchange interactions. This means that the polarization vector $\bm{P}$ is generally not aligned with the propagation direction of the pumping beam. However, when the optical pumping rate is much larger than $\Gamma_{1,0}$ and $\Gamma_{2,0}$, atomic polarization is relatively large and the direction of $\bm{P}$ is close to the pumping beam direction, which is the high-polarization limit.

\subsection{Heading error analysis}~\label{sec:head}
For ground state alkali atoms with a nuclear spin of $I$ in a bias field of $\mathbf{B}$, its Hamiltonian is expressed as
\begin{equation}
H=A_{hf}\bm{I}\cdot\bm{S}+g_s\mu_B\bm{S}\cdot\mathbf{B}-g_I\mu_B\bm{I}\cdot\mathbf{B}.
\end{equation}
Here, $A_{hf}$ is the ground state hyperfine constant, $\bm{S}$ is the electron spin, $\mu_B$ is the Bohr magneton, and the nuclear magnetic moment is $\bm{\mu}_I=g_I\mu_B\bm{I}$ as in Ref.~\cite{lee2021}, which leads to an opposite sign of $g_I$ compared with that in Refs.~\cite{Arimondo1977,steckrb87}. The resulted energy shift for non-stretched states, relative to the original level without considering the hyperfine and Zeeman effect, can be extracted from the Breit-Rabi formula~\cite{breit1931,seltzerthesis}
\begin{equation}~\label{eq:Ens}
E_{ns}=-\frac{\hbar\omega_{hf}}{2(2I+1)}-g_Im_F\mu_B\mathrm{B}\pm\frac{\hbar\omega_{hf}}{2}\sqrt{1+\frac{4xm_F}{2I+1}+x^2},
\end{equation}
with $\hbar\omega_{hf}$ as the hyperfine splitting between ground states, $x=(g_s+g_I)\mu_B\mathrm{B}/\hbar\omega_{hf}$, and $\pm$ refers to $F=I\pm1/2$. The energies of two stretched states contain only linear Zeeman terms with the expression as
\begin{equation}~\label{eq:Es}
E_{s}=\frac{I}{2I+1}\hbar\omega_{hf}\pm\frac{g_s-2Ig_I}{2}\mu_B\mathrm{B},
\end{equation}
with $\pm$ refering to $m_F=\pm(I+1/2)$. By keeping the Zeeman interaction terms up to the second order, we get the Zeeman splitting between any pair of $|F, m_F\rangle$ and $|F, m_F-1\rangle$ from Eqs.~\eqref{eq:Ens} and \eqref{eq:Es} as
\begin{equation}~\label{eq:dE}
\Delta E=(-g_I\mu_B\pm\mu_{eff})\mathrm{B}\mp\frac{\mu_{eff}^2\mathrm{B}^2}{\hbar\omega_{hf}}(2m_F-1),
\end{equation}
with $\mu_{eff}=(g_s+g_I)\mu_B/(2I+1)$ as the effective atomic magnetic dipole moment.

The first term on the right side of Eq.~\eqref{eq:dE} corresponds to the aforementioned LZE, and it introduces an energy difference of 2$g_I\mu_B$B between the Larmor frequency of atom spin in the upper ground hyperfine state and that in the lower one. The second term corresponds to the NLZE. For an optically pumped magnetometer based on $^{87}$Rb atoms in upper ground hyperfine states and the high-polarization limit as discussed in the previous subsection, we follow the treatment in Ref.~\cite{lee2021} by assuming a spin temperature atomic population distribution, and the measured field with the NLZE induced heading error is related with the measured Larmor frequency $\omega_L$ as
\begin{equation}~\label{eq:B}
\mathrm{B}_{87}\approx\frac{4\hbar\omega_L}{(g_s-3g_I)\mu_B}\left[1+\frac{3\omega_L}{\omega_{hf}}\cos\theta\frac{P(7+P^2)}{5+3P^2}\right].
\end{equation}
{Please note that the equation above is different from the one in Ref.~\cite{lee2021} because $\theta$ is defined differently.} The result extracted from $^{85}$Rb atoms treated in the same approximation is:
\begin{equation}~\label{eq:B85}
\mathrm{B}_{85}\approx\frac{6\hbar\omega_L}{(g_s-5g_I)\mu_B}\left[1+\frac{3\omega_L}{\omega_{hf}}\cos\theta\frac{P(P^4+18P^2+21)}{3P^4+14P^2+7}\right].
\end{equation}

\section{Experiment setup and measurement scheme}~\label{sec:setup}
\begin{figure}[htb]
\includegraphics[width=3in]{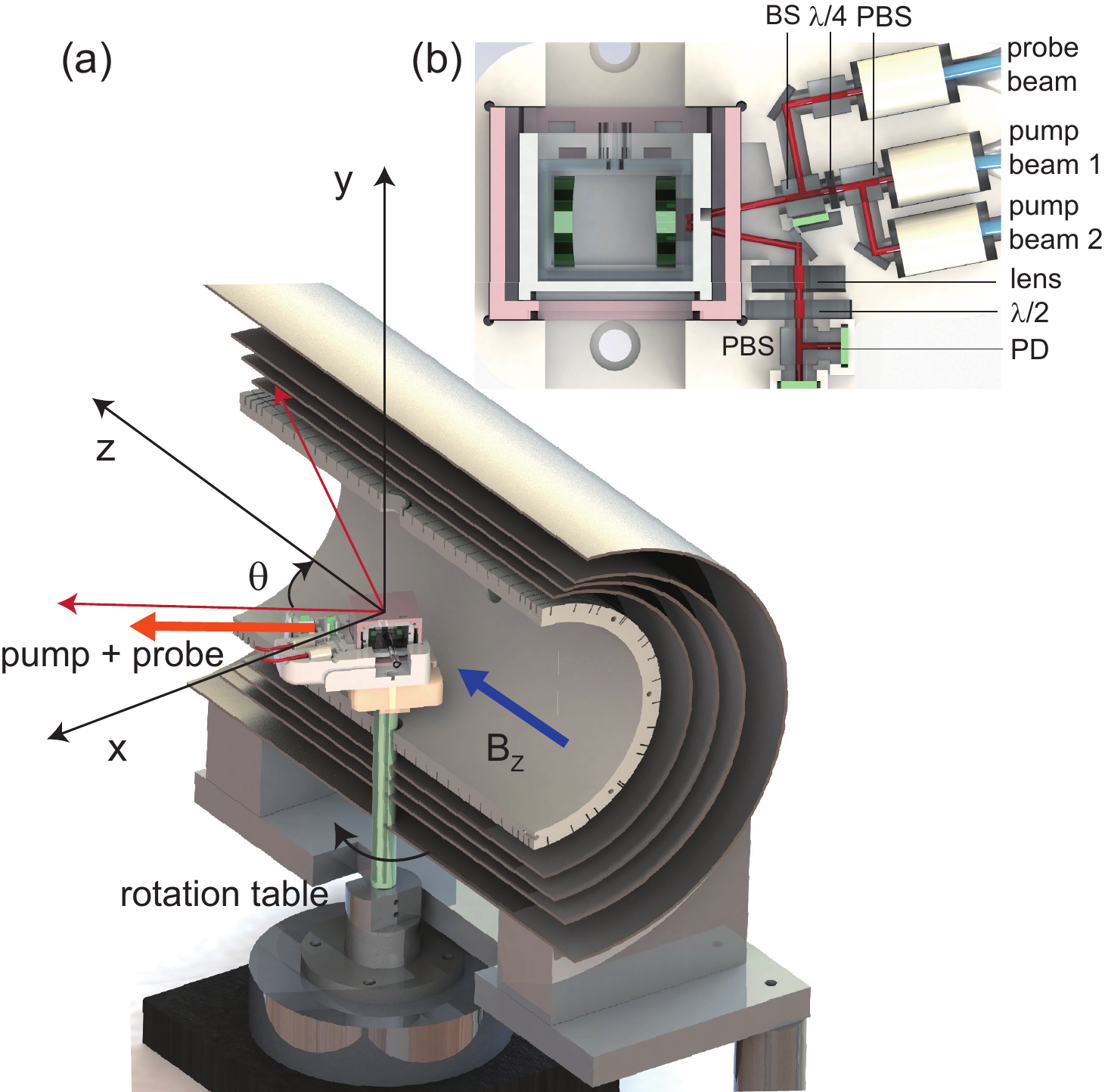}
\caption{\label{fig:setup} Illustration of the experiment setup, with the inset as the top view of the 3D-printed optical platform for the scalar FID magnetometer.}
\end{figure}

The magnetometer sensor sits in the middle of five-layer mu-metal shields, with a bias field $\mathrm{B}_z$ generated by solenoid coils inside the shields, as shown in Fig.~\ref{fig:setup}(a). In order to minimize the magnetic field gradients, we have compensated field gradient components of $\partial \mathrm{B}_z/\partial z$, $\partial^2 \mathrm{B}_z/\partial z^2$, and $\partial \mathrm{B}_z/\partial x$. In the end, the field gradient of $\mathrm{B}_z$ along $x$ and $z$ directions are both less than 0.1 nT/mm. The sensor is also connected with a high-precision rotation table outside the shields through an epoxy holder, so that the relative orientation between the pumping beam and the bias field direction can be precisely controlled. 


Figure~\ref{fig:setup}(b) shows the optical platform of the magnetometer, which is manufactured by high-precision 3D printing technology. To improve the signal-to-noise ratio of the magnetometer, a Herriott-cavity-assisted vapor cell~\cite{liu2023} is used to increase the interaction length between light and atoms. This Herriott cavity consists of two identical cylindrical mirrors~\cite{Silver2005}, with a radius of curvature of 60 mm, a diameter of 12.7 mm, a thickness of 2.5 mm, a separation of 11.5 mm and a relative angle between symmetrical axes of 52.2$^\circ$. Such a cavity is bonded on a piece of silicon wafer and attached to the glass cell using the anodic bonding technique~\cite{cai2020}. {As discussed in Ref.~\cite{liu2023}, the direction of the beam that passes through the cavity is defined by the averaged beam orientation inside the cavity, which is along the connection of both cavity mirror centers.} The atomic cell is filled with N$_2$ gas of 400 Torr and a droplet of $^{87}$Rb atoms. This cell is normally operated at a temperature around 75$^\circ$C by running high-frequency ac currents through ceramic heaters. 

The pumping and probe beams are independently generated by distributed-Bragg-reflector (DBR) laser diodes, and coupled to the sensor by polarization-maintaining fibers. To identify and suppress certain sources of heading error, we couple two pumping beams on resonance with Rb D1 transition into the sensor. These two beams are combined by a polarizing beam-splitter (PBS) so that they have orthogonal polarizations afterwards. They are circularly polarized after passing through a quarter-wave plate ($\lambda/4$), and combined with the probe beam, which is 54 GHz blue detuned from the Rb D1 line, by a non-polarizing beam-splitter before entering the atomic cell. The transmitted probe beam is analyzed by a polarimeter, which consists of a {half}-wave plate ($\lambda/2$), a PBS and two photodiode detectors (PD).

\begin{figure}[htb]
\includegraphics[width=3in]{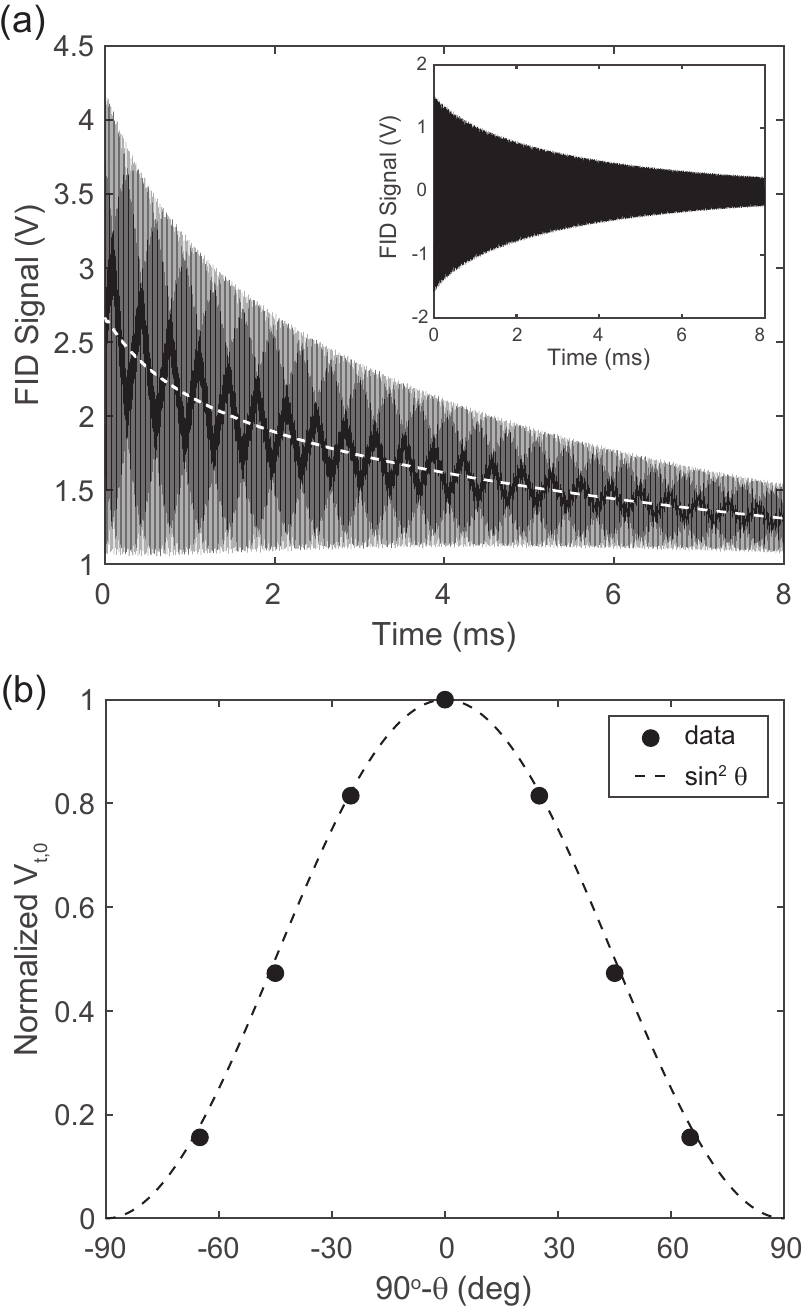}
\caption{\label{fig:FID_signal} (a) Measured FID signals at B=10 $\mu$T and $\theta$=45$^\circ$, with the white dash line showing the signal after a low-pass filter and the inset showing the signal after a high-pass filter. (b) The initial amplitude of transverse signals ($V_{t,0}$) as a function of $\theta$, with the dotted line showing the $\sin^2{\theta}$ relation.}
\end{figure}

The magnetometer operates in a pulsed pump-probe mode with a period of 16 ms, {and the bandwidth of this magnetometer is 31.25 Hz limited by the Nyquist sampling theorem}. In the first half period, one of the pumping beam is turned on with its power modulated at a frequency close to the atomic Larmor frequency by a fiber acoustic-optical modulator with a duty cycle of 20\%. The time-averaged pumping beam power before entering the cell is 1.13 mW. In the next half period, the pumping beam is turned off, and the FID signal of atomic polarization is recorded. During the whole period, the probe beam is kept on with an input beam power of 0.7 mW. A typical signal is shown in Fig.~\ref{fig:FID_signal}~(a), which includes contributions from both transverse ($V_t(t)$) and longitudinal ($V_l(t)$) polarizations:
\begin{eqnarray}
\label{eq:h3}
V(t)&=&V_t(t)+V_l(t)+V_0\nonumber\\
&=&V_{t,0}\sin(\omega_L t+\phi)e^{{-t}/{T_2}}+V_{l,0}e^{{-t}/{T_1}}+V_0,
\end{eqnarray}
where $V_{t,0}$ and $V_{l,0}$ corresponds to the initial {amplitude} of the transverse and longitudinal signal, respectively. The free parameters $T_1$ and $T_2$ represent the decay time of the longitudinal and transverse atomic polarization, and $V_0$ represents the offset voltage from the electronics. We can also separate the signals from these two polarization components by adding a high-pass or low-pass filter to the signal. {In addition, we have also checked the atomic diffusion effects by assigning two free parameters to the transverse polarization decay time~\cite{Li2011}, and found no difference in the extracted $\omega_L$. Therefore, we neglect the atomic diffusion effect in this work.}

In the experiment, we focus on the transverse part of the FID signal, and get the fitting results of $\omega_L$. The magnetic field magnitude $\mathrm{B}$ is initially extracted from $\omega_L$ using the linear term in Eq.~\eqref{eq:dE}, 
\begin{equation}~\label{eq:B1}
\mathrm{B}\approx\hbar\omega_L/(\mu_{eff}-g_I\mu_B).
\end{equation}
For an arbitrary $\theta$, the atomic transverse polarization can be expressed as $P_t(t)=P_0\sin\theta\cos(\omega t)e^{-t/T_2}$ in the high-polarization limit, where $P_0=\sqrt{P_{t,0}^2+P_{l,0}^2}$ as shown in Eqs.~\eqref{eq:Pz} and ~\eqref{eq:Pt}. The recorded transverse signal $V_t(t)$ is proportional to the projection of the transverse polarization on the probe beam, which is $V_t(t)\propto P_t(t)\sin\theta=P_0\sin^2\theta\cos(\omega t)e^{-t/T_2}$. Therefore, the initial amplitude of the transverse signal ($V_{t,0}(\theta)$) follows a $\sin^2(\theta)$ function in the high-polarization limit, which is confirmed by the results in Fig.~\ref{fig:FID_signal}(b). This indicates a detection dead zone when $\theta$ is close to zero, and the measurement in this work is limited in the range of $25^\circ\leq\theta\leq 155^\circ$. 

Two independent methods are developed to calibrate the atomic polarization $P$ in this system. The first one is to extract the polarization from the pumping beam transmission. In the high atomic polarization limit, if the input beam has an intensity of $I_i$ and a photon spin of $\boldsymbol{s}$ ($|s|=1$), the transmitted beam intensity $I_t$ is dependent on $P$ according to the relation
\begin{equation}
\label{eq:h4}
I_t=I_{i}e^{-\mathrm{OD}|s-P|},
\end{equation}
where OD is the optical depth on resonance.  The second method is to vary the input pumping beam power, measure the corresponding $V_{t,0}$ and $V_{l,0}$ in  Eq.~{\eqref{eq:h3}}, and extract the atomic polarization by fitting the results {of these initial amplitudes} using Eqs.~\eqref{eq:Pz} and ~\eqref{eq:Pt}. Atomic polarizations calibrated by the above two ways are denoted as $P_1$ and $P_2$, respectively. For a $^{87}$Rb atomic magnetometer working at 75 $^\circ$C, the results of $P_1$ and $P_2$ are listed in Tab.~\ref{tab:P}.  It can be concluded that, for the {experimental} conditions in this work, the results from both calibrations agree with each other within 5\%, and are all in the high polarization regime, which is consistent with the initial assumption. In addition, the difference of the field magnitude calculated from Eq.~\eqref{eq:B} using $P_1$ or $P_2$, is on the order of 0.1~nT, which is negligible.

\begin{table}
\caption{Atomic polarizations calibrated by two different methods ($P_1$ and $P_2$ defined in the main text) for a $^{87}$Rb atomic magnetometer working at 75 $^\circ$C pumped by an input beam power with a 20\% modulation duty cycle and a time-averaged power of 1.13 mW.}
\label{tab:P}
  \begin{tabular}{|c | c| c | c| c| }
  \hline
  $|90^\circ-\theta|$&0$^\circ$&25$^\circ$&45$^\circ$&65$^\circ$\\
  \hline
  $|P_1|$&0.92&0.93&0.95&0.98\\
  \hline
  $|P_2|$&0.84&0.87&0.92&0.94\\
  \hline
\end{tabular}
\end{table}

\section{Identifying and suppressing heading errors}

{In this paper, the statistical error of the magnetometer measurements is well below 1 nT, which is neglected in the rest of the paper. The measurement uncertainty is dominated by heading errors discussed in the following parts.}

\subsection{LZE-induced heading error}

We measure the sensor heading error by comparing the fitted result B($\theta$) from Eq.~\eqref{eq:B1} at different sensor orientations. The two main sources of heading errors are LZE and NLZE, mentioned in Sec.~\ref{sec:head}. According to Eqs.~\eqref{eq:dE} and \eqref{eq:B}, LZE-induced heading error is independent of the sign of atomic polarization, while it is the opposite case for the NLZE-induced heading error. Therefore, we can separate these two effects by comparing the results with opposite atomic orientations. 

\begin{figure}[htb]
\includegraphics[width=3in]{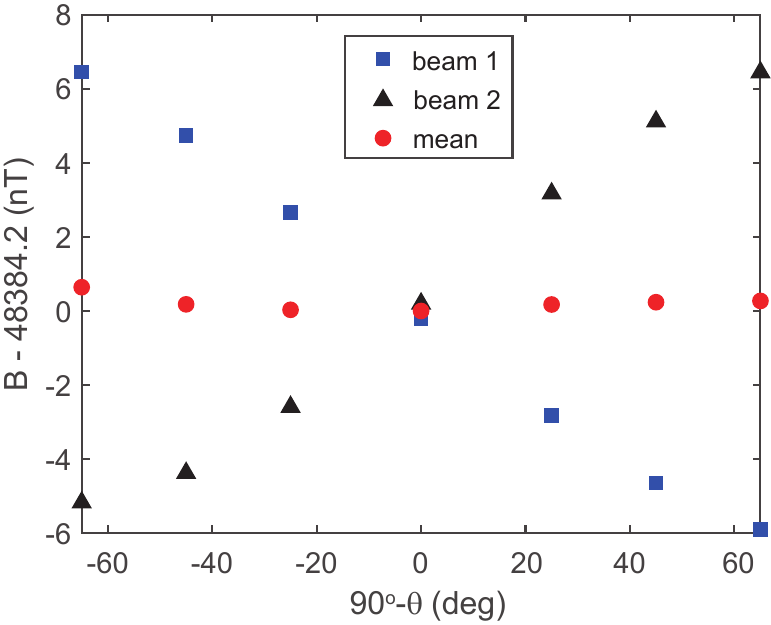}
\caption{\label{fig:comp} Experiment results of heading errors using opposite circularly polarized pumping beams at a bias field around 50~$\mu$T.}
\end{figure}

Figure~\ref{fig:comp} shows the measured field magnitude as a function of the sensor orientation when the bias field is around 50 $\mu$T. We eliminate the NLZE induced heading error by taking the average of the measured results with pumping beams 1 and 2 in Fig.~\ref{fig:setup}, which have opposite circular polarizations. The remaining heading error, which is negligible compared with the NLZE-induced heading error, is mainly contributed by LZE and the residual field $\mathrm{B}_{res}$ from the sensor head. $\mathrm{B}_{res}(\theta)$ is independently calibrated by measuring the heading error {at 2.5 $\mu$T and 5~$\mu$T, and the results at these two fields agree with each other}. From these measurements, we can conclude that the LZE-induced heading error is within 0.5 nT in the measured range of $\theta$.

\begin{figure}[htb]
\includegraphics[width=3in]{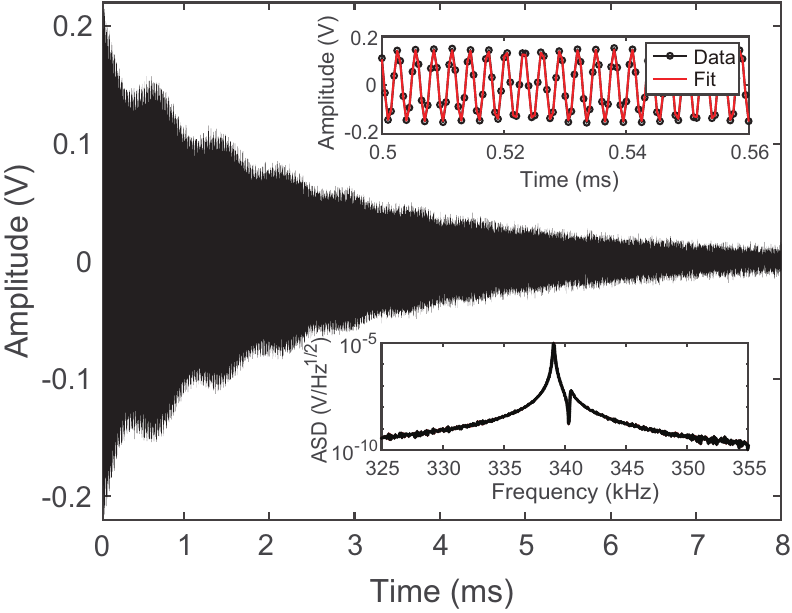}
\caption{\label{fig:LZE} Experiment results of $^{87}$Rb FID signals at B=50~$\mu$T, with the pumping-beam power modulation frequency close to the Larmor frequency of $F=1$ ground states, which is about 1.4 kHz above the Larmor frequency of $F=2$ ground states. The upper inset shows the fitting result using Eq.~\eqref{eq:h3}, while the lower inset shows the amplitude spectral density (ASD) of the measurement.}
\end{figure}

While the pumping beam modulation frequency $\omega_m$ is normally set at the Larmor frequency of the upper hyperfine state ($\hbar\omega_m\approx(\mu_{eff}-g_I\mu_B)B$) to maximize the signal, we also test the extreme case where $\omega_m$ is set at the Larmor frequency of the lower hyperfine state ($\hbar\omega_m\approx(\mu_{eff}+g_I\mu_B)B$), with the data shown in Fig.~\ref{fig:LZE}. The beating signal in the time domain comes from the difference in linear Zeeman shifts (2$g_I\mu_B$B) between two hyperfine states as shown in Eq.~\eqref{eq:dE}. We have tried to use two different models to fit the data in Fig.~\ref{fig:LZE}, one is Eq.~\eqref{eq:h3} with a single frequency parameter and the other one is a double-frequency fitting model. The frequency extracted from the single-frequency model agrees with the lower-frequency parameter extracted with the double-frequency model within 1 Hz. This is due to the fact that, even in this extreme case, atoms are still dominantly distributed at the upper hyperfine state which has a unique dark state for optical pumping.


\subsection{NLZE-induced heading error and its suppression scheme}

\begin{figure}[htb]
\includegraphics[width=3in]{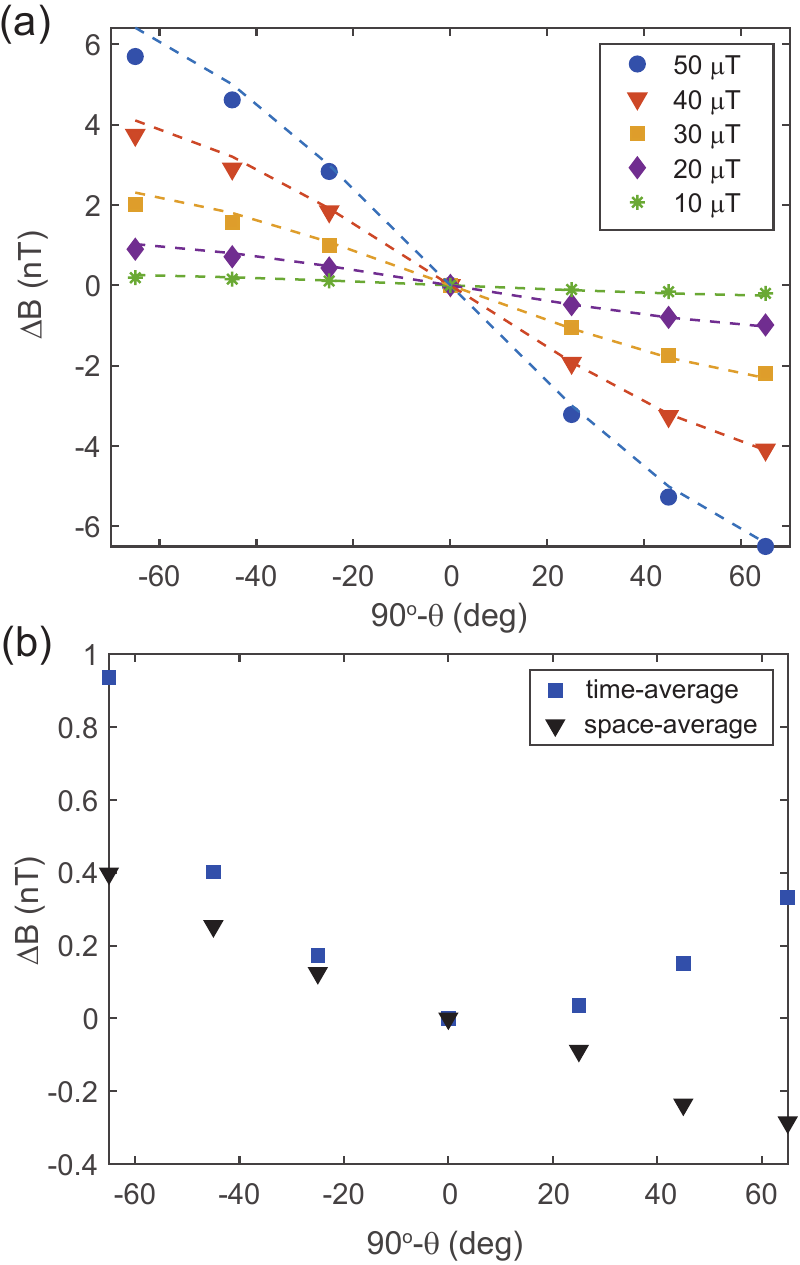}
\caption{\label{fig:heading_mea_sup} (a) Measured heading errors as a function of $\theta$ in the field range of 10 $\sim$ 50 $\mu$T (solid markers) and theoretical prediction results using Eq.~\eqref{eq:B} (dashed lines). (b) The remaining heading error after averaging out atomic longitudinal polarization in time or space.}
\end{figure}

In the rest of the paper, we use the parameter $\Delta \mathrm{B}$ to characterize heading errors with the residual field effect $\mathrm{B}_{res}$ removed,
\begin{equation}
\Delta \mathrm{B}=\mathrm{B}(\theta)-\mathrm{B}(\theta=90^\circ)-\mathrm{B}_{res}(\theta).
\end{equation}
The measurement results of $\Delta \mathrm{B}$ of a $^{87}$Rb magnetometer as a function of $\theta$ in the field range of 10 to 50 $\mu$T are shown in the Fig.~\ref{fig:heading_mea_sup}(a). The measured experiment data points agree well with the predictions using Eq.~\eqref{eq:B} and calibrated atomic polarizations, which also confirms that atom populations follow the spin-temperature distribution in a magnetometer using a Bell-Bloom optical pumping method. 

In the following, we focus on suppressing the heading error either directly or indirectly. Here, the direct methods refer to suppressing the heading error without using the measurement results, while the indirect methods suppress the heading error by using a comagnetometer system and combining multiple atomic precession frequencies, which are the most direct and accurate parameters extracted from the FID magnetometer. 

For the direct methods, we try to suppress the NLZE-induced heading error by eliminating the horizontal atomic polarization, as indicated in Eq.~\eqref{eq:B}. This can be achieved by averaging out the horizontal atomic polarization either in time or space. In the former case, both pumping beam 1 and 2 in Fig.~\ref{fig:setup} are used, where the pumping beam 2 is turned on with the same modulation frequency and duty cycle as beam 1, except a $\pi$ phase delay. In practice, the ratio of two beam powers are adjusted so that $V_{l,0}$ in Eq.~\eqref{eq:h3} is kept nulled. In this way, the transverse atomic polarization is built up constructively by the two pumping beams while the longitudinal atomic polarization are averaged out. The squared data points in Fig.~\ref{fig:heading_mea_sup}(b) confirm that this scheme helps to suppress the heading error within 1 nT over the sensor orientations used in this work. In free-space experiments, this scheme can also be realized by modulating the pumping beam polarization with an electro-optic modulator.

To average out the longitudinal atomic polarization in space, a half-wave plate is added in the middle of the Herriott cavity~\cite{yu2022}. In this way, the pumping beam polarization is flipped each time it passes the wave plate, and the measurement result is effectively the average of the results from two atomic ensembles pumped by beams with opposite polarizations. The triangle data points in Fig.~\ref{fig:heading_mea_sup}(b) show that this space-average scheme achieves similar performance on suppressing the heading error as the time-average one. It needs to be noted that, in practice, the quality of the pumping beam polarization degrades as it reflects on the mirror surfaces and passes through the wave plate, which may contribute to the imperfect suppression result.

To demonstrate the indirect methods to suppress the NLZE-induced heading error, we make use of a Rb isotope comagnetometer. Suppose that the precession frequencies of both Rb isotopes can be simultaneously measured using the same setup, then the same bias field can be independently expressed by the precession frequency of each isotope as
\begin{equation}~\label{eq:B2}
\mathrm{B}\approx a_i\omega_{L,i}+b(P)_i\omega^2_{L,i}\cos\theta. 
\end{equation}
Here, $i$ denotes the isotope $^i$Rb, and $a$ and $b$ represents the coefficients for linear and nonlinear terms in Eqs.~\eqref{eq:B} and \eqref{eq:B85}. The results of two isotopes can be combined to cancel the NLZE terms with the extracted field value as
\begin{equation}~\label{eq:Bcomb}
\mathrm{B}=\frac{c(P)a_{87}\omega_{L,87}-a_{85}\omega_{L,85}}{c(P)-1},
\end{equation}
where $c(P)=b(P)_{85}\omega^2_{L,85}/b(P)_{87}\omega^2_{L,87}=\Delta \mathrm{B}_{85}/\Delta \mathrm{B}_{87}$. 

\begin{figure}[htb]
\includegraphics[width=3in]{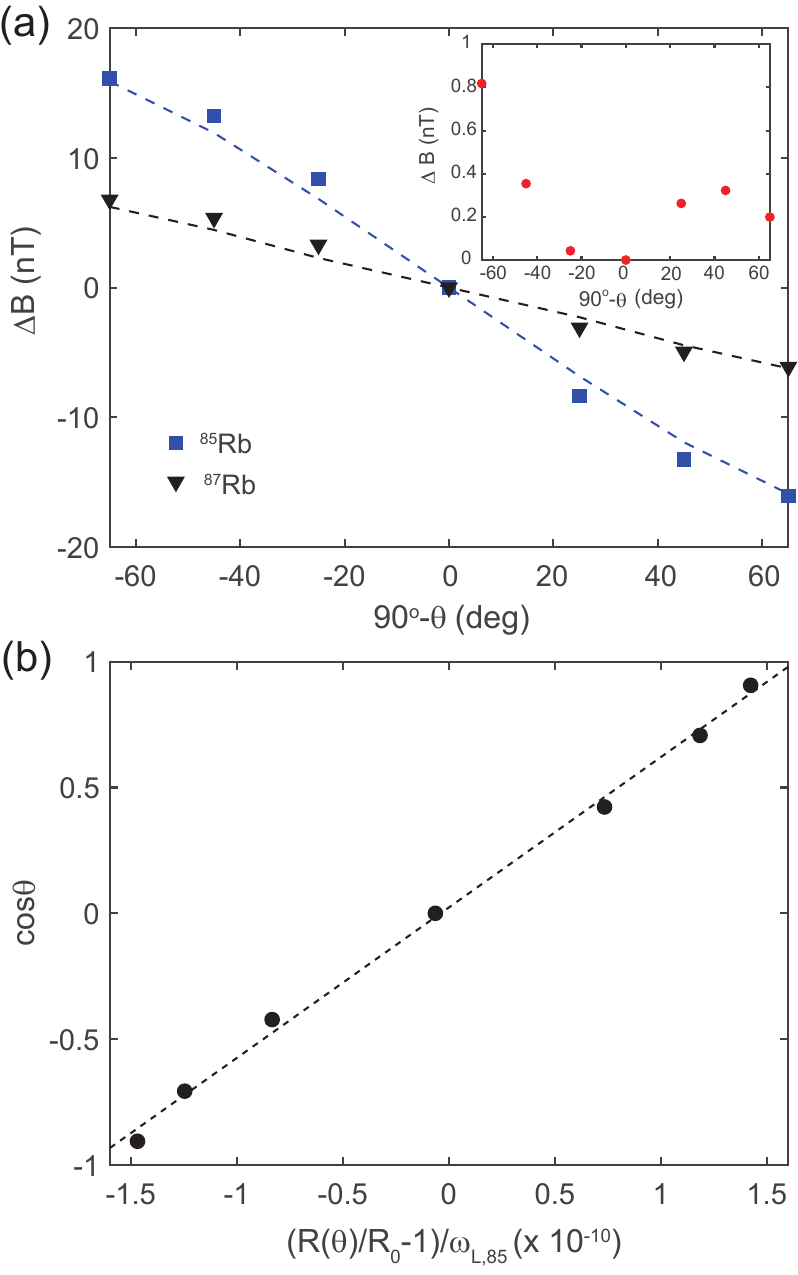}
\caption{\label{fig:8587_R}(a) Experiment results of heading errors simultaneously measured by two Rb isotopes at B=50~$\mu$T, compared with theoretical predictions based on Eqs.~\eqref{eq:B} and ~\eqref{eq:B85}. The inset shows the remaining heading error when the value of $c(P)$ at $\theta=25^\circ$ ($\bar{c}$=2.57) is used is used in Eq.~\eqref{eq:Bcomb}. (b) Experiment results of $(R(\theta)/R_0-1)/\omega_{L,85}$ at different $\cos\theta$ under the same experiment conditions as plot~(a), where the dash line shows a linear fitting with a coefficient of determination of 0.997.}
\end{figure}

This scheme is implemented by using a cell consisting of Rb atoms with natural abundance, while all other conditions are same as the previous $^{87}$Rb cell. Due to the pressure broadening effect, the same pumping and probe beams work well for both isotopes. The pumping beam is amplitude modulated at a frequency of $\omega_{t}$, which is approximately equal to $\omega_{L,87}/3$ and  $\omega_{L,85}/2$, and the modulation duty cycle is 10\%. As shown in Tab.~\ref{tab:pdual}, the atomic polarization of $^{85}$Rb is larger than $^{87}$Rb due to a larger effective duty cycle for the isotope with a smaller gyromagnetic ratio. In addition, the atomic polarization decreases as $\theta$ gets close to 90$^\circ$, and the minimum atomic polarization is below 0.75 for both isotopes at $\theta=90^\circ$. Figure~\ref{fig:8587_R}(a) shows the comparison between experimental data for each isotope and theoretical predictions based on Eqs.~\eqref{eq:B} and \eqref{eq:B85}. Here the discrepancy between them when $\cos\theta$ is small may be attributed to the relatively low atomic polarization in these cases where the feasibility of Eqs.~\eqref{eq:B} and \eqref{eq:B85} decrease. 

\begin{table}
\caption{Absolute values of atomic polarizations and $c(P)$ for simultaneously optical pumping both Rb isotopes.}
\label{tab:pdual}
  \begin{tabular}{|c | c| c | c| c| }
  \hline
  $|90^\circ-\theta|$&0$^\circ$&25$^\circ$&45$^\circ$&65$^\circ$\\
  \hline
  $^{85}$Rb polarization&0.72&0.76&0.84&0.93\\
  \hline
  $^{87}$Rb polarization&0.57&0.63&0.79&0.92\\
  \hline
  $c(P)$&3.16&3.01&2.70&2.57\\
  \hline
\end{tabular}
\end{table}

Since the polarization $P$ changes with $\theta$ in the experiment, the value of $c(P)$ is actually depending on $\theta$, as listed in Tab.~\ref{tab:pdual}. When the value of $\theta$ is unknown, we need to use a constant $\bar{c}$ as the approximation of $c(P)$ in Eq.~\eqref{eq:Bcomb} to minimize the NLZE for the whole experiment range of $\theta$. In practice, we find that $\bar{c}$ can be chosen near the value of $c(P)$ at $\theta$ farthest away from 90$^\circ$. For example, in our experiment, the heading error can be suppressed within 1 nT when $\bar{c}$ is chosen in the range of 2.40 to 2.62, and a case with $\bar{c}$ equal to $c(P)$ at $\theta$=25$^\circ$ ($\bar{c}$= 2.57) is demonstrated in the inset of Fig.~\ref{fig:8587_R}(a). 

The common readout from the Rb-isotope comagnetometer is the frequency ratio $R=\omega_{L,87}/\omega_{L,85}$. According to Eq.~\eqref{eq:B2}, $R$ is dependent on $\theta$, and in the high-polarization limit, this relation can be expressed as
\begin{eqnarray}~\label{eq:theta_w}
\cos\theta&=&\frac{a_{85}}{b(P)_{85}-R^2(\theta)b(P)_{87}}\frac{R(\theta)/R_0-1}{\omega_{L,85}}\nonumber\\
&=&k(P)\frac{R(\theta)/R_0-1}{\omega_{L,85}},
\end{eqnarray}
where $R_0=a_{85}/a_{87}=3(g_s-3g_{I,87})/2(g_s-5g_{I,85})=1.49886$. It should be noted that $k(P)$ is independent of the bias field magnitude, and its value at the high-polarization limit is 6.37 $\times 10^9$ s$^{-1}$. 

Figure~\ref{fig:8587_R}(b) shows the experiment results of $(R(\theta)/R_0-1)/\omega_{L,85}$ with $\cos\theta$ at a bias field around 50 $\mu$T. The experiment results can be well fitted by a linear relation, which is due to the high atomic polarization at relatively large values of $\cos\theta$ as shown in Tab.~\ref{tab:pdual}, and the slope is extracted from the linear fitting as $(5.98\pm0.15)\times 10^9$ s$^{-1}$, which agrees with the fore-mentioned high-polarization limit value of $k(P)$. In practice, we can extract the value of $\cos\theta$ using the calibrated $k(P)$ and measured frequency ratio $R(\theta)$, and correct the heading error using Eqs.~\eqref{eq:B} and ~\eqref{eq:B85}. Compared with the way to extract $\cos\theta$ using the amplitude information of longitudinal and transverse signals in Ref.~\cite{lee2021}, the method here relies only on the frequency information.

\section{Conclusion}

In conclusion, we have studied different sources of heading errors in a Rb FID magnetometer. By employing the Bell-Bloom optical pumping method, the heading error induced by the nuclear spin LZE is almost eliminated,  and the NLZE-induced error is identified as the main source in this system. To address this problem, we have developed several different suppression methods, which have been demonstrated to keep the heading error under 1 nT at a bias field of 50 $\mu$T over a sensor orientation range of $25^\circ\leq\theta\leq 155^\circ$. The schemes based on a single magnetometer rely on averaging out the atomic longitudinal polarization, and have the advantages of maintaining the signal magnitude and correcting the heading error automatically. The schemes based on a comagnetometer do not require additional hardware costs in the sensor head, and only need the information of the recorded precession frequencies. In the following work, we plan to continue the efforts in this paper and focus on understanding and reducing the residual heading error. We also hope to apply the scalar magnetometers studied in this work to precision measurements. Though preliminary results for this application are promising~\cite{wang2024}, special attentions are required for systematic effects besides heading errors~\cite{abel2020,rosner2022}. Moreover, we want to further develop a vector atomic magnetometer.

\section{Acknowledgments}
This work was partially carried out at the University of Science and Technology of China (USTC) Center for Micro and Nanoscale Research and Fabrication. This work is supported by National Science Foundation of China (Grant No. 12174372).

\end{document}